\pdfoutput=1  

\documentclass[11pt,a4paper]{article}

\usepackage[utf8]{inputenc}
\usepackage[T1]{fontenc}
\usepackage{amsmath,amssymb}
\usepackage{graphicx}
\usepackage[margin=2.5cm]{geometry}
\usepackage[numbers,sort&compress]{natbib}
\usepackage[colorlinks=true,linkcolor=blue,citecolor=blue,urlcolor=blue]{hyperref}

\title{On the Stability of the Euler--Poisson Dark-Fluid Model}

\author{Bal\'azs Endre Szigeti$^{1,2}$\thanks{Corresponding author: \texttt{szigeti.balazs@wigner.hu}, ORCID: \href{https://orcid.org/0000-0002-8028-962X}{0000-0002-8028-962X}}
\and Imre Ferenc Barna$^{1}$\thanks{\texttt{barna.imre@wigner.hu}, ORCID: \href{https://orcid.org/0000-0001-6206-3910}{0000-0001-6206-3910}}
\and Gergely G\'abor Barnaf\"oldi$^{1}$\thanks{\texttt{barnafoldi.gergely@wigner.hu}, ORCID: \href{https://orcid.org/0000-0001-9223-6480}{0000-0001-9223-6480}}}

\date{%
$^{1}$\,Wigner Research Centre for Physics, Institute for Particle and Nuclear Physics, Budapest HU-1121, Hungary\\
$^{2}$\,E\"otv\"os Lor\'and University, Faculty of Informatics, Department of Algorithms and their Application, Budapest HU-1117, Hungary\\[2ex]
\today}

\begin{document}

\maketitle

\begin{abstract}
We present a stability analysis of a dark-fluid model described as a non-relativistic, rotating, non-viscous, self-gravitating fluid. We assume spherical symmetry and model the matter by a polytropic equation of state. The resulting coupled nonlinear partial differential equation system is solved using a self-similar ansatz known from the Guderley\,--\,Landau\,--\,Stanyukovich problem. We find that three of the four Lyapunov exponents are negative, while only one is positive. This indicates one unstable direction and three contracting directions in phase space. The single positive exponent is relatively small, suggesting that the self-similar solution is only weakly unstable and remains dynamically robust over the investigated interval.
\end{abstract}


\section{Introduction}
By assuming various symmetries, one can successfully solve a wide variety of dynamical systems described by nonlinear partial differential equations (PDEs). One particularly common and useful type of symmetry is self-similarity. Numerous physically relevant self-similar solutions have been found since Gottfried Guderley’s famous discovery of spherically symmetric self-similar solutions for an imploding gas collapsing toward the center~\cite{gouderley}. In this paper, we use the self-similar solutions introduced independently by Leonid Ivanovich Sedov and Sir Geoffrey Ingram Taylor during the 1940s~\cite{sedov,ref-Taylor}.

Although such models have been well known for decades, they continue to attract considerable attention. This \emph{ansatz} has already been applied successfully in several contexts, including heat conduction~\cite{imre1}, the three-dimensional Navier--Stokes and Euler equations~\cite{imre4}, and star formation~\cite{ref-Guo}. The concept of self-similarity also has a wide range of applications in general relativity. Homothetic solutions were first introduced by Cahill and Taub~\cite{ref-cahill} and have been studied extensively in connection with gravitational collapse~\cite{ref-collapse} and asymptotic cosmological solutions~\cite{ref-cosmology}.

This perspective is particularly relevant in astrophysical and cosmological settings, where some of the most fundamental open problems concern the dark sector. One possible approach is the dark-fluid concept, which seeks to describe both dark matter and dark energy within a single continuous classical medium~\cite{ref-Arbey}.
In our previous studies, we developed and investigated a cosmological model based on hydrodynamics~\cite{Barna(2022)}. The dynamics of the dark-fluid-like medium are governed by a coupled nonlinear system of partial differential equations. In our model, we studied one of the simplest dark fluid materials described by a linear equation of state (EoS). The Euler equation governs fluid dynamics, whereas the associated gravitational field is determined through the Poisson equation. We found time-dependent scaling solutions for the velocity, density, and gravitational fields that could serve as candidates for describing the evolution of gravitationally coupled dust-like dark matter within a Newtonian cosmological framework. Our present goal is to give a stability analysis of our model and derive the Lyapunov exponents.

\section{The Model}
\label{themodel}

We consider a set of coupled non-linear partial differential equations, which describe the non-relativistic dynamics of a compressible, self-gravitating, rotating fluid with zero thermal conductivity and zero viscosity, in radial spherical coordinates.

The first governing equation is the continuity equation. The second is the Euler equation, which includes the pressure term derived from the EoS, the radial component of the external force density, and an effective rotational term on the right-hand side. Here, $\omega$ is a dimensionless parameter describing the magnitude of rotation, and $\theta$ is the polar angle. We assume that the rotation is sufficiently slow that a spherically symmetric leading-order description remains applicable, i.e., the rotational energy is negligible compared with the gravitational energy. The third equation is the Poisson equation for the gravitational field.
\begin{subequations}
\begin{align}
    \partial_t \rho + (\partial_r \rho) u + (\partial_r u) \rho + \dfrac{2u\rho}{r} &= 0 \label{eq:02A} \ , \\
    \partial_t u + (u\partial_r)u &= - \dfrac{1}{\rho} \partial_r P  - \partial_r \Phi + \dfrac{\sin \theta \omega^2r}{t^2}, \   \\
    2 \partial_r \Phi  + r\partial_{rr}\Phi &= 4 \pi G r \rho.
\end{align}
\label{eq:02}
\end{subequations}

Here, the dynamical variables are $\rho = \rho(r,t)$, $u = u(r,t)$, $\Phi = \Phi(r,t)$, and $P = P(r,t)$, denoting the density, radial velocity, gravitational potential, and pressure, respectively. We apply the linear EoS
\begin{equation}
    P(\rho) = w \rho^n, \quad \quad n = 1 \ .
\label{eq:03}
\end{equation}
More information about the model can be found in~\cite{Szigeti2023}. Several forms of the EoS are available in astrophysics, and polytropic ones have been used successfully in the past; see, for example, Emden's classical book~\cite{ref-Emden}. Equations of state with negative pressure of this type are widely used in dark-fluid cosmology, most notably in the Chaplygin and generalised Chaplygin gas models \cite{Kamenshchik2001,Bento(2002),Gorini2003}. In Eq.~\eqref{eq:03}, the parameter $w$ may vary depending on the type of matter governing the system's evolution. Traditionally, $w=0$ corresponds to the EoS of ordinary non-relativistic matter or cold dust. In this paper, we choose $w=-1$ as a simple phenomenological case within the present dark-fluid framework. The adiabatic speed of sound can be evaluated from Eq.~\eqref{eq:03}, and it is straightforward to show that it is constant. Note that, in the calculations below, the geometrized unit system ($c = 1$, $G=1$) is used.

One can find semi-analytic solutions of the equations by applying the long-established self-similar \emph{ansatz} by Sedov and Taylor~\cite{sedov,ref-Taylor}, which can be expressed in the following form
\begin{subequations}
\begin{gather}
    u (r,t)  = t^{-\alpha} f \bigg( \dfrac{r}{t^{\beta}} \bigg) \ ,
    \label{eq::SedovTaylorAnsatzA}\\
    \rho(r,t) = t^{-\gamma} g \bigg( \dfrac{r}{t^{\beta}} \bigg) \ , \label{eq::SedovTaylorAnsatzB}\\
    \Phi(r,t) = t^{-\delta} h\bigg( \dfrac{r}{t^{\beta}} \bigg), \label{eq::SedovTaylorAnsatzC}
\end{gather}
\label{eq::SedovTaylorAnsatz}
\end{subequations}
where $t$ denotes time and $r$ the radial coordinate. One can see that all shape functions $(f,g,h)$ depend only on the combination $rt^{-\beta}$; therefore, we introduce the self-similarity variable $\zeta$. The variable $\zeta$ is dimensionless in geometrized units. The exponents $\alpha$, $\beta$, $\gamma$, and $\delta$ are called similarity exponents and have clear physical meaning. In particular, $\beta$ characterises the temporal spreading of the spatial profile when $\beta>0$, or contraction when $\beta<0$. The remaining exponents describe the temporal scaling of the amplitudes of the corresponding fields. The similarity density profile \(g(\eta)\) is restricted by the physical admissibility condition \(g(\eta)\geq 0\).

A general description of the properties of these types of scaling solutions can be found in our previous publication~\cite{Barna(2022)}. Thus, we have calculated the relevant time and space derivatives of the shape functions and substituted them into the equations~\eqref{eq:02}.  We obtained the following numerical value for the exponents $\alpha = 0$, $\beta = 1$, $\gamma = 2$, and $\delta = 0$ for both the non-rotating and the rotating cases. These exponent values indicate a spreading spatial profile for the relevant dynamical variables. In this sense, the self-similar behaviour may be interpreted as expansion-like at large astrophysical or cosmological scales.

\begin{figure}[!ht]
\centering
\includegraphics[width=0.55\textwidth]{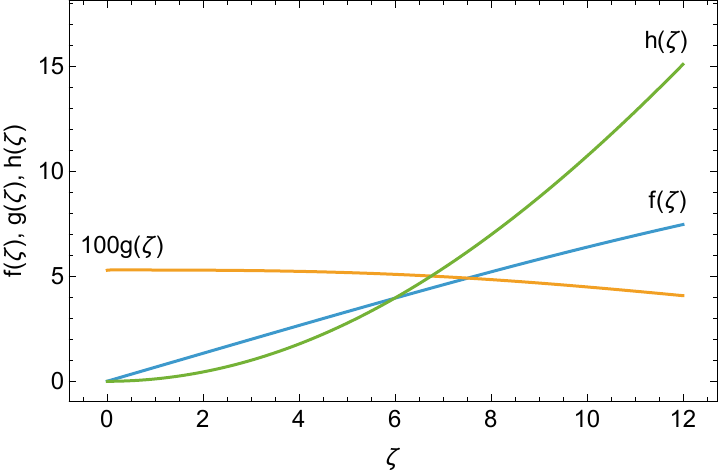}
\caption{Numerical solutions of the shape functions (spherical symmetric case, linear EoS), the integration was started at $\zeta_{0} = 0.001$, and the initial conditions of $f(\zeta_{0}) = 0.05$, $g(\zeta_{0}) = 0.053$, $h(\zeta_{0}) = 0$, and $h'(\zeta_{0}) = 1$ were used. For better visibility, the function $g(\zeta)$ was scaled up by a factor of 100. The values are expressed in geometrized units.
\label{fig1}}
\end{figure}

By substituting the obtained numerical values of the similarity exponents, we have reduced the induced PDE system into an ordinary differential equation (ODE) system that depends only on the $\zeta$ independent variable. We found that the obtained equation system has the following form,
\begin{subequations}
\begin{align}
    \zeta [ f(\zeta) g(\zeta) ]' + 2f(\zeta)g(\zeta) &=  2 g(\zeta) +\zeta^2 g'(\zeta), \label{eq::11A}\\
    -\zeta f'(\zeta) + f'(\zeta) f(\zeta) & = - \dfrac{w g'(\zeta)}{g(\zeta)}  - h'(\zeta)  + \zeta \omega^2\sin\theta  ,\label{eq::11B}\\
    2h'(\zeta) + h''(\zeta) \zeta & =  4 \pi g (\zeta) \zeta . \label{eq::11C}
\end{align}
\label{eq::11}
\end{subequations}
Unfortunately, the presented ordinary differential equation system Eq.~\eqref{eq::11} cannot be solved analytically. For linearised non-autonomous ordinary differential equation systems, the stationary point of the phase space can be found, and one can say something about the general asymptotic behaviour of the solutions~\cite{Kenneth(2018)}. Nonetheless, there is no generally known method for non-linearised non-autonomous differential equation systems. Local existence for the Euler--Poisson Cauchy problem has been studied in the mathematical literature, including density profiles without compact support \cite{BrauerKarp2015}. Here we restrict attention to a self-similar solution class, defined on the time interval where the similarity ansatz is regular. Therefore, it is a reasonable approach to solve the obtained ordinary differential equation system, Eqs.~\eqref{eq::11}, numerically for a large number of parameter sets (based on physical considerations) to explore the behaviour of the solution of the system with different boundary and initial conditions. One example of the numerical solution is seen in Fig.~\ref{fig1}.

Here at a specific parameter set and initial conditions, the velocity shape function $f(\zeta)$ is nearly linear after a short initial decrease. The function $g(\zeta)$ becomes asymptotically flat after a rapid rise, which is consistent with matter conservation. The final shape function, $h(\zeta)$, exhibits an increasing polynomial trend associated with the gravitational potential. To obtain sufficiently smooth numerical solutions, we solved the ODE system using the adaptive numerical integrator provided by \emph{Wolfram Mathematica 13.1}~\cite{ref-Wolfram}. For all calculations, the integration limits were $\zeta_{0}=0.001$ and $\zeta_{\max}=40$, as in Ref.~\cite{Szigeti2023}. We considered the following ranges of initial conditions: $f(\zeta_0)=0.005$-$0.5$, $g(\zeta_0)=0.001$-$0.1$, together with $h(\zeta_0)=0$ and $h'(\zeta_0)=1$.

This choice of initial conditions reflects the physically reasonable assumption that the density profile satisfies $\mathcal{R}(g) \subset \mathbb{R}^{+}$ and remains finite. Some recent works suggest that dark fluid may have negative mass~\cite{Farnes(2018)}. However, in the present model, this choice leads to singular solutions. Likewise, $\mathcal{R}(f) \subset \mathbb{R}^{+}$ corresponds to an initially radially expanding fluid. We observed numerically that if the initial values of $f(\zeta)$ and $g(\zeta)$ are chosen outside the ranges given above, the solution becomes singular. We also found that varying the initial condition associated with the gravitational potential does not qualitatively affect the time evolution of the system, but only produces vertical shifts. Therefore, we set its initial value to zero.

\section{The stability of the self-similar solution}

The Sedov\,--\,von Neumann\,--\,Taylor self-similar \emph{ansatz} for the Euler\,--\,Poisson system provides a powerful framework for analysing non-linear gravitating fluid dynamics. By introducing an appropriate similarity variable, the original partial differential equations governing velocity, density, and gravitational potential can be reduced to a coupled system of ordinary differential equations. Even though this reduction substantially simplifies the mathematical structure of the original problem, it typically leads to a dynamical system that is explicitly dependent on the similarity variable, rendering it inherently non-autonomous.

The non-autonomous nature of Sedov-type self-similar Euler\,--\,Poisson equations has important consequences for stability analysis. Classical approaches based on autonomous fixed-point theory~\citep{Agarwal2001} or asymptotic~\citet{Lyapunov(1992)} exponents rely on infinite-time limits and time-translation invariance, assumptions that are generally violated in self-similar configurations. In particular, the evolution variable in the reduced system does not represent physical time, and the dynamics often terminate at finite values corresponding to singularity formation, shock emergence, or loss of self-similarity~\citep{Barenblatt(1996)}. As a result, stability must be understood in a finite-interval sense, along dynamically evolving solutions.

In this context, finite-time Lyapunov methods offer a natural and mathematically consistent approach to stability analysis. By quantifying the growth or decay of infinitesimal perturbations along a reference self-similar trajectory over a prescribed interval of the similarity variable, finite-time Lyapunov exponents capture transient instabilities and local sensitivity that are intrinsic to non-autonomous gravitational flows. This trajectory-based perspective is particularly well-suited to self-similar Euler\,--\,Poisson dynamics.

Motivated by these considerations, the stability properties of the self-similar Euler\,--\,Poisson system studied in this work are analyzed using finite-time Lyapunov exponents computed along the obtained reference solutions. This framework provides a quantitative measure of dynamical stability that remains valid in the absence of autonomous structure, allowing for a systematic investigation of transient growth phenomena relevant to gravitational collapse and self-similar flow evolution. Thus, one should consider a general non-autonomous dynamical system of the form
\begin{equation} \label{eq:ode_system_ftle}
\frac{\mathrm{d}\mathbf{x}}{\mathrm{d}\zeta} = \mathbf{F}(\mathbf{x}, \zeta),
\qquad
\mathbf{x} \in \mathbb{R}^n \text{ and } \zeta \in \mathbb{R}^{+},
\end{equation}
where $\mathbf{F}: \mathbb{R}^n \times \mathbb{R} \to \mathbb{R}^n$ is assumed to be sufficiently smooth to ensure the existence and uniqueness of solutions~\citep{coddington1955}. Let $\mathbf{x}(\zeta; \zeta_0, \mathbf{x}_0)$ denote the solution satisfying the initial condition $\mathbf{x}(\zeta_0)=\mathbf{x}_0$. The corresponding flow, $\varphi$ is defined as
\begin{equation}
\varphi_{\zeta_0}^{\zeta}(\mathbf{x}_0) = \mathbf{x}(\zeta; \zeta_0, \mathbf{x}_0).
\end{equation}
For non-autonomous systems, the flow satisfies, for any intermediate point $\tau$
\begin{equation}
    \varphi^{\zeta}_{\zeta_0}(\zeta; \zeta_0,\mathbf{x}_0)=\varphi^{\zeta}_{\tau}(\zeta; \tau,\varphi^{\tau}_{\zeta_0}(\tau; \zeta_0,\mathbf{x}_0)).
\end{equation}
Throughout this analysis, a single reference trajectory is selected by specifying $\mathbf{x}_0$ and integrating the system over a finite interval $[\zeta_0,\zeta_f]$. Let us consider a small perturbation $\delta(\zeta)$ from the reference trajectory $\mathbf{x}(\zeta)$, such as
\begin{equation}
    \mathbf{x}(\zeta) \to  \mathbf{x}(\zeta) + \delta \mathbf{x}(\zeta).
\end{equation}
The evolution of the perturbation is governed by the variational equation
\begin{equation}
\frac{\mathrm{d}}{\mathrm{d}\zeta}\,\delta\mathbf{x} = D_{\mathbf{x}}\mathbf{F}\bigl(\mathbf{x}(\zeta),\zeta\bigr)\,
\delta\mathbf{x},
\label{eq:variational_eq}
\end{equation}
where $D_{\mathbf{x}}\mathbf{F}$ denotes the Jacobian matrix of the vector field evaluated along the reference solution. The solution of the differential equation, defined in Eq.~\eqref{eq:variational_eq}, can be expressed in terms of the fundamental matrix
\begin{equation}
\mathbf{\Phi}({\zeta_0},{\zeta}) = D\varphi_{\zeta_0}^{\zeta}(\zeta; \zeta_0,\mathbf{x}_0),
\end{equation}
which maps initial perturbations to their evolved values,
\begin{equation}
\delta\mathbf{x}(\zeta) = \mathbf{\Phi}(\zeta,\zeta_0)\, \delta\mathbf{x}(\zeta_0),
\end{equation}
with the following properties
\begin{align}
    \mathbf{\Phi}(\zeta,\tau)\mathbf{\Phi}(\tau,\zeta_0) & = \mathbf{\Phi}(\zeta, \zeta_0) \ , \label{eq::semigroup_property} \\
    \mathbf{\Phi}(\zeta,\tau)& =\mathbf{\Phi}^{-1}(\tau,\zeta) \ ,\\
    \mathbf{\Phi}(\tau, \tau) &= \hat I \ .
\end{align}
For a given initial perturbation direction $\delta \mathbf{x}_0$, the finite-time Lyapunov exponent can be defined as
\begin{equation}
    \lambda(\zeta, \zeta_0, \delta \mathbf{x}_0) = \dfrac{1}{\zeta-\zeta_0} \ln \dfrac{\|\mathbf{\Phi}(\zeta,\zeta_0)\, \delta \mathbf{x}_0 \|}{\| \delta \mathbf{x}_0\|},
\end{equation}
over the interval $[\zeta_0,\zeta_f]$, where $\|\cdot \|$ is the $L^2$-norm. Yet, Eq.~\eqref{eq:variational_eq} cannot be solved analytically for most real-world physical systems. In numerical implementations, it is often advantageous to avoid the explicit construction of $\mathbf{\Phi}({\zeta_0},{\zeta_f})$. Instead, the finite-time Lyapunov exponents can be obtained via the algorithm of~\citet{Benettin1980}

\subsection{Benettin\,--\,Galgani\,--\,Giorgilli\,--\,Strelcyn algorithm}

The algorithm divides the total integration range $[\zeta_0, \zeta_f]$ into $N$ subintervals of length $\delta \zeta = (\zeta_f - \zeta_0)/N$. The validity of this partition of intervals is the direct consequence of the semigroup property (cf. Eq.~\eqref{eq::semigroup_property}) of the fundamental matrix \citep{Benettin1980}, such as
\begin{equation}
\mathbf{\Phi}(\zeta_f, \zeta_0) = \mathbf{\Phi}(\zeta_N, \zeta_{N-1}) \mathbf{\Phi}(\zeta_{N-1}, \zeta_{N-2}) \cdots \mathbf{\Phi}(\zeta_1, \zeta_0).
\end{equation}
At every step, the algorithm employs a periodic Gram\,--\,Schmidt re-orthonormalisation to prevent numerical overflow and maintain linear independence of perturbation vectors~\cite{Gram1883,Schmidt1907}. We implemented this re-orthonormalisation step here via the QR decomposition~\citep{Francis(1961)},
\begin{equation} \label{eq::qr_decomposition}
Y^{(k)} = \mathbf{Q}^{(k)}\mathbf{R}^{(k)},
\end{equation}
which is equivalent to Gram\,--\,Schmidt orthogonalisation in a numerically robust matrix formulation. The \(Q^{(k)}\) is an orthogonal matrix and \(R^{(k)}\) is an upper triangular matrix with positive diagonal entries. Here \(Y^{(k)}\) denotes the matrix whose columns are the evolved perturbation vectors over the \(k\)-th integration interval.

The diagonal entries \(R^{(k)}_{ii}\) measure the stretching of the \(i\)-th reorthonormalised perturbation vector during this interval. They are therefore local growth factors generated by the QR reorthonormalisation procedure. Consequently, in the Benettin--Galgani--Giorgilli--Strelcyn algorithm, the Lyapunov characteristic exponents are estimated from the accumulated logarithmic QR growth factors,
\begin{equation} \label{eq::lyapunov_estimates}
    \lambda_i^{\mathrm{QR}}
    =
    \frac{1}{\zeta_f-\zeta_0}
    \sum_{k=1}^{N}
    \ln \left| R^{(k)}_{ii} \right| .
\end{equation}
These quantities are QR-based finite-interval estimates associated with the chosen reference trajectory and reorthonormalisation procedure. They should be distinguished from finite-time Lyapunov exponents defined directly through the singular values of the full fundamental matrix,
\begin{equation}
    \lambda_i^{\mathrm{FTLE}}
    =
    \frac{1}{\zeta_f-\zeta_0}
    \ln \sigma_i\!\left(\Phi(\zeta_f,\zeta_0)\right).
\end{equation}
In the asymptotic limit, under the usual convergence assumptions, the QR-based Benettin estimates converge to the Lyapunov spectrum.

\section{Results}
\label{sec:res}
To analyse the stability properties of the obtained solution, the spherically symmetric Euler\,--\,Poisson equations (cf. Eq.~\eqref{eq::11}) must be transformed into the form defined in Eq.~\eqref{eq:ode_system_ftle}. Consequently, we introduce a new auxiliary function $k(\zeta) = h'(\zeta)$, and the relevant ODE system reduces to first order, as follows
\begin{subequations}
\begin{align}
    f'(\zeta) & = \frac{\left(\zeta -f(\zeta )\right) \left(2w-\zeta  k(\zeta) + \zeta^2\omega^2\sin\theta\right)}{\zeta  \left[ \left(f(\zeta) -\zeta \right)^2 -w \right]},\\
    g'(\zeta) & =g(\zeta )  \frac{-2 \left( f(\zeta ) -\zeta \right)^2 +\zeta  \left(k(\zeta )-\omega^2\sin\theta  \right)}{\zeta  \left[ \left(f(\zeta) -\zeta \right)^2 -w \right]}, \\
    k'(\zeta) & = 4 \pi  g(\zeta )-2k(\zeta )\zeta^{-1} , \\
    h'(\zeta) & = k(\zeta).
\end{align}
\label{eq::explicit_ode_system}
\end{subequations}
with the effective rotational term. The vector field $\mathbf{F}$, whose components are defined as the r.h.s. of Eqs.~\eqref{eq::explicit_ode_system}, exhibits singularities where the denominators vanish. Thus, the initial conditions established in Ref.~\cite{Szigeti2023} were used for the finite-interval stability analysis. To avoid numerical instabilities from numerical differentiation, the Jacobian (cf. Eq.~\eqref{eq:variational_eq}) is obtained symbolically, as follows
\begin{equation} \label{eq::Jacobian}
D_{\mathbf{x}}\mathbf{F}(\mathbf{x}(\zeta), \zeta) =
\left(
\begin{array}{cccc}
 J_{11} & 0 &  J_{13} & 0 \\
 J_{21} & J_{22} & J_{23} & 0 \\
 0 & 4 \pi  & -\frac{2}{\zeta } & 0 \\
 0 & 0 & 1 & 0 \\
\end{array}
\right),
\end{equation}
{\footnotesize
\begin{align*}
    J_{11} & = \frac{1}{\zeta}\frac{\left[ \left(f(\zeta) -\zeta \right)^2 +w \right]^2 }{\left[ \left(f(\zeta) -\zeta \right)^2 -w \right]^2}
    \left[2 w-\zeta  k(\zeta )+ \zeta^2\omega^2 \right],\\
    J_{13} & =  \frac{f(\zeta ) - \zeta }{\left[ \left(f(\zeta) -\zeta \right)^2 -w \right]}, \\
    J_{21} & = \frac{g(\zeta)}{\zeta^2} \frac{ 2 \left( f(\zeta ) - \zeta  \right)^2  \left[ 2 \left(f(\zeta ) -\zeta  \right)^2+3 w\right]- \left[ \zeta k(\zeta ) - \zeta^2 \omega^2\right]\left[3 \zeta ^2+3 f^2(\zeta )-2 \zeta f(\zeta) )+w\right]}{\left[ \left(f(\zeta) -\zeta \right)^2 -w \right]^3} \left[\zeta  k(\zeta )+\zeta^2\omega^2-2 w \right], \\
    J_{22} & = \frac{1}{\zeta^2}\frac{\left(f(\zeta )- \zeta \right) \left[2 \left(f(\zeta )-\zeta \right)^2-\zeta  k(\zeta )+\zeta^2\omega^2\right] }{ \left[ \left(f(\zeta) -\zeta \right)^2 -w \right]^2} \left[\zeta  k(\zeta )-\zeta^2\omega^2-2 w\right], \\
    J_{23} &= 2 \frac{g(\zeta )}{\zeta}\frac{\left(f(\zeta ) -\zeta \right)  }{ \left[ \left(f(\zeta) -\zeta \right)^2 -w \right]^2}\left[ \left(f(\zeta) -\zeta\right)^2-\zeta  k(\zeta )+\zeta^2\omega^2+w\right],
\end{align*}
}
where $\sin\theta =1$. One can explicitly state the numerical iteration. The interval $[\zeta_0,\zeta_f]$ is divided into $N$ subintervals,
\begin{equation}
\zeta_k=\zeta_0+k\Delta\zeta, \qquad \Delta\zeta=\frac{\zeta_f-\zeta_0}{N}, \qquad k=0,\ldots,N.
\end{equation}
The initial perturbation basis is chosen as \(Q_0=I_n\). On the \(k\)-th subinterval, the variational equation
\begin{equation}
\frac{dY_k}{d\zeta}
=
D_xF\bigl(x(\zeta),\zeta\bigr)Y_k,
\qquad
Y_k(\zeta_{k-1})=Q_{k-1},
\end{equation}
is integrated simultaneously with the reference trajectory, using the Jacobian defined in Eq.~\eqref{eq::Jacobian}. At
\(\zeta_k\), the evolved perturbation matrix is factorised as
\begin{equation}
Y_k(\zeta_k)=Q_kR_k,
\end{equation}
where \(Q_k\) is orthogonal and \(R_k\) is upper triangular. The matrix \(Q_k\) is used as the initial perturbation basis on the next subinterval, while the logarithmic stretching factors
\begin{equation}
        S_{i,k}
    =
    S_{i,k-1}
    +
    \ln\left|(R_k)_{ii}\right|.
\end{equation}
are accumulated from the diagonal elements of \(R_k\).  The resulting QR-based finite-interval Lyapunov estimates are defined in Eq.~\eqref{eq::lyapunov_estimates}, which can be written as a function of $S_{i,k}$
\begin{equation}
\lambda_i^{\mathrm{QR}} =
    \frac{S_{i,N}}{\zeta_f-\zeta_0}.
\end{equation}
The finite-interval Lyapunov exponents are obtained as follows
\begin{equation}
    \lambda_1 = 0.241, \quad  \lambda_2=-0.649, \quad \lambda_3=-1.241, \quad \lambda_4 = -2.701.
\end{equation}
If a small rotational term is added to the system defined in Eqs.~\eqref{eq::explicit_ode_system} and the Jacobian in Eq.~\eqref{eq::Jacobian} is modified accordingly, the Lyapunov exponents vary to
\begin{equation}
    \lambda_1 = 0.278, \quad  \lambda_2=-0.757, \quad \lambda_3=-1.418, \quad \lambda_4 = -3.19.
\end{equation}
Both spectra exhibit a single positive Lyapunov exponent, indicating the existence of one unstable mode. The remaining three negative exponents indicate contraction along three directions of the local tangent dynamics over the investigated interval. Consequently, the obtained solution may be interpreted as a weakly unstable self-similar solution that remains dynamically robust in a local finite-interval sense, although it requires fine-tuning of one parameter to satisfy the physical matching conditions.

\section{Conclusions and Outlook}

In this work, we investigated the stability of self-similar solutions of the Euler\,--\,Poisson dark-fluid model. By applying a Sedov\,--\,Taylor-type self-similar \emph{ansatz}, we reduced the original nonlinear partial differential equation system to a set of ordinary differential equations. These equations were solved numerically, and the resulting solutions were analysed using finite-time Lyapunov exponents.

We found that, for the parameter ranges considered, three of the four Lyapunov exponents are negative, while one is positive. This indicates that the self-similar solution possesses a single unstable mode and three contracting directions in phase space. The same qualitative behaviour remains present when a small rotational term is included. These results suggest that the model admits a weakly unstable but dynamically robust self-similar solution.

\section*{Author contributions}
Conceptualisation, I.F.B.; formal analysis, B.E.Sz.; software, B.E.Sz.; visualisation, B.E.Sz.; writing---original draft preparation, B.E.Sz.; writing---review and editing, I.F.B. and G.G.B. All authors have read and agreed to the published version of the manuscript.

\section*{Funding}
Authors gratefully acknowledge the financial support by the Hungarian National Research, Development and Innovation Office (NKFIH) ADVANCED\_25 K153456, 2024-1.2.5-T\'ET-2024-00022, 2025-1.1.5-NEMZ\_KI-2025-00005 (GGB), and the COST Action FuSe (CA24101). Authors are grateful for the possibility to use the GenAI4Science service of the HUN-REN Cloud (see H\'eder et al. 2022; \url{https://science-cloud.hu/}) which helped us achieve the results published in this paper.

\section*{Data availability}
This work is based primarily on analytic derivations and numerical calculations. The data underlying this article, including the numerical outputs used to generate the plots, will be shared on reasonable request to the corresponding author.

\section*{Acknowledgments}
Authors gratefully acknowledge the useful discussion with Bal\'azs P\'al.

\section*{Conflicts of interest}
The authors declare no conflict of interest.

\bibliographystyle{unsrtnat}
\bibliography{Reference}

\end{document}